# DESIGN METHODOLOGY AND CONSIDERATIONS FOR NOVA 53 MHZ RF CAVITIES*

C. Ader[#] and D.W. Wildman
Fermi National Accelerator Laboratory, Batavia, IL 60510, USA

*Abstract*

The NOvA Experiment will construct a detector optimized for electron neutrino detection in the existing Neutrino at Main Injector (NuMI) beamline. This beamline is capable of operating at 400 kW of primary beam power and the upgrade will allow up to 700 kW. The cavities will operate at 53 MHz and three of them will be installed in the Recycler beamline. Thermal stability of the cavities is crucial since this affects the tuning. Results of finite element thermal and structural analysis involving the copper RF cavity will be presented.

## BACKGROUND

The NOvA project will convert part of the Fermilab Recycler Ring to a pre-injector to the Main Injector, allowing the cycle time to be reduced from two seconds to 1.33 seconds. This will yield an increase in beam power and double the intensity of the NuMI beam line by accelerating six Booster batches in the Main Injector with twice as many protons per batch, which is possible via slip stacking.

Three new 53 MHz cavities will be built for Slip Stacking in the Recycler. These cavities are of the quarter wave resonator type and have a much lower R/Q compared with the MI cavities for beam stability and reduced beam loading voltages. The new cavities will be able to deliver a peak voltage of 150 KV and will have enough cooling to run CW [1]. Two RF cavities with different frequencies are used for the 12-batch slip stacking. Six batches are injected on the central frequency of the first RF system, 52.809 MHz, and captured with 110 kV RF voltage, then deaccelerated. Cavity and beam-line parameters are given in Table 1.

Table 1: 53 MHz cavity and beam-line parameters.

| Beam Energy | 120 GeV/c |
|---|---|
| Protons per Pulse | $4.90 \times 10^{13}$ |
| Total Beam Injected, protons | $5.16 \times 10^{13}$ |
| Cycle Time, sec | 1.33 |
| Beam Sigma, mm (rms) in (x,y) plane | 1.3 |
| Pulse Length, sec | $10 \times 10^{-6}$ |
| Peak Gap Voltage, kV | 150 |
| Frequency, MHz | 52.809 |
| Slipping Frequency Difference, Hz | 1,260 |

The second RF cavity is off during the first six batch injections. After the frequency of the first cavity is lowered, the seventh batch is injected on the central frequency of the second RF system and captured with 110 kV RF voltage, then accelerated. This continues until the twelfth batch, and when the bunches are aligned, the beam is extracted from the Recycler and injected into the Main Injector where they are captured in RF buckets corresponding to a 2 MV RF voltage. Therefore, during the next turn, the MI will accelerate six Booster batches that have twice as many protons. This creates a more efficient accelerator because the cycle in the MI is decreased from two seconds to 1.33 seconds since slip stacking is performed in the Recycler ring [2].

## RF DESIGN

A solid model of the desired cavity shown in Fig. 1 existed in IDEAS and it was faster to import this geometry into ANSYS Workbench, rather than creating the geometry internally. Although some geometry manipulation within the Finite Element Analysis is generally necessary, it is best to keep this to a mininum. Importation may be completed via a transfer format such as IGES, STEP or a Parasolid. In general, the solid model should be simplified as much as possible before transferring to ANSYS which would include removing unneeded components. Also, the model should make maximum use of symmetry conditions [3].

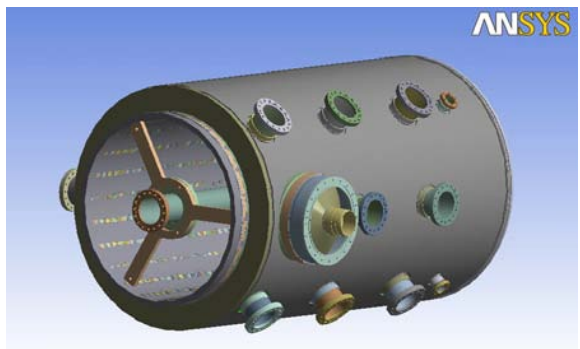

Figure 1: Solid model of the 53 MHz cavity.

Most conventional RF cavities are fabricated from copper or aluminum. Aluminum presents some difficulties due to poor vacuum characteristics. Aluminum tends to absorb water leading to oxidation that increases the coefficient of secondary electron emission, a phenomena that is unacceptable for storage rings.

The combined electrical, mechanical and vacuum characteristics of copper make it a better choice for this application. However, the relatively large physical size, the large quantity of access ports needed, and the anticipated difficulties in joining similar or dissimilar materials to the main body were obstacles that had to be overcome. Oxygen-free copper (OFHC) was selected because of the purity of the copper.

## FINITE ELEMENT ANALYSIS

The finite element (FE) 53 MHz cavity model generated using ANSYS Workbench [4] calculates temperature profile and corresponding directional displacement due to thermal strain. In general for the cavity assembly, solid 70 and solid 90 (thermal elements) as well as solid 186 and solid 187 (structural elements) were used to construct the model.

### Thermal Modeling

As part of the physics design and analysis of a cavity, the overall operating parameters and performance characteristics are established. RF power fed into the cavity is continuous; therefore a steady-state analysis was applied. The peak power in the cavity is 150 kW and half of this is dissipated in the outer cavity while the other half in the inner cavity according to this mathematical relationship in Eq. 1:

$$Power \propto \cos^2(\frac{\pi}{2} \times \frac{x}{L}) \qquad (1)$$

The variables x and L are defined in Fig. 2.

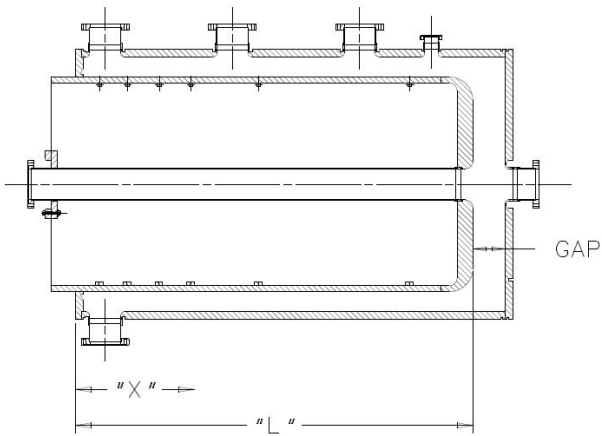

Figure 2: Elevation view of the 53 MHz cavity.

The inner cavity was cut into nine equal length, 13.4 cm bins in order to apply the appropriate heat load to each bin and then the number of water cooling lines was varied in order to come up with an optimal number. As a baseline calculation for the 75 kW design, 4 tubes per bin were applied. The power in each bin is shown in Fig. 3. Low Conductivity Water (LCW) flow, delivered at a temperature of 30 ± 2°C removes 75 kW of energy from each cavity conductor, inner and outer. The inner cavity cooling system consists of thirty-six parallel 19 mm x 19 mm copper tube. The ϕ10 mm inner diameter tube supports a velocity of 1.5 m/sec and water heat transfer film coefficient of 9,550 (W/m²-°C).

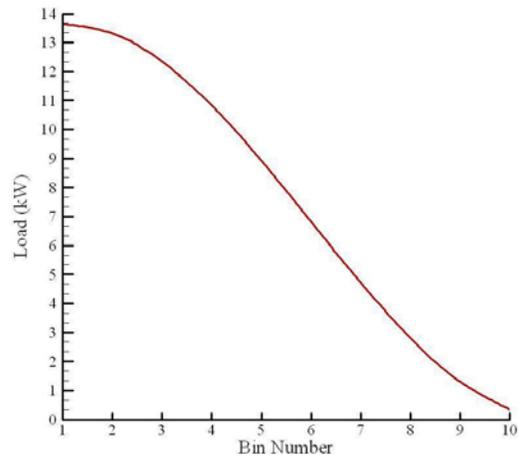

Figure 3: Cavity power load as a function of bin number.

### Thermal Results

Under steady-state conditions, the inner cavity shown in Fig. 4 remained beneath the 10°C design differential with a maximum temperature of 41.9°C at the upstream end.

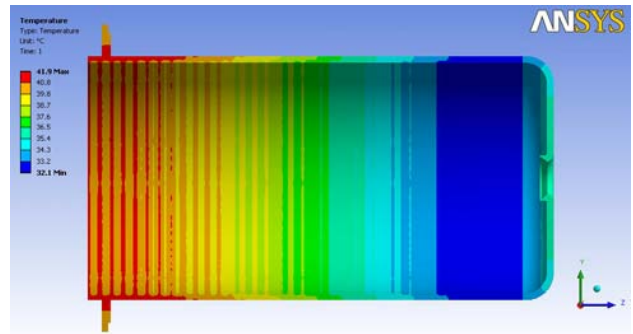

Figure 4: Temperature profile of the inner cavity.

### Stress and Pressure Modeling

In ANSYS, the thermal elements are converted to structural elements in order to obtain the stress and displacement solutions. The thermal distortion of the cavity is based on the coefficients of thermal expansion of the cavity materials and the nodal temperature data obtained from the thermal solution, which is applied as a load on the structural model. Other loads applied to the model include acceleration due to gravity and external atmospheric pressure. Also, symmetry boundary conditions and cavity support constraints were applied.

The primary consideration with the cavity is the vacuum load that is exerted on the cavities. There is atmospheric pressure on the inside of the conductor cavity and ultra-high vacuum on the outside of the inner conductor and the inside of the outer cavity and also on the inside of the vacuum pipe where beam passes through.

The hoop stress of the cavity is 13.9 MPa and the longitudinal stress is 3.6 MPa. Using Section UG-28 ASME Pressure Vessel Code, the allowable working stress was calculated to be 0.99 MPa which is greater than the pressure seen by the cavity. The collapse pressure is calculated to be 0.35 MPa which is greater than the atmospheric pressure of 0.101 MPa, so therefore the wall thickness is adequate.

The vacuum load on the large end plate must be considered since this affects the gap distance. A deflection of 0.95 mm of deflection is acceptable if the tolerance is ± 0.635 mm and is repeatable [5]. In these calculations, the end plate is assumed to be 19 mm thick although actually it is now 25.4 mm thick. Roark and Young [6] calculate 0.69 mm deflection while the ANSYS Model determines it to be 0.86 mm deflection. The boundary conditions include gravity, vacuum loads, and one fixed and one sliding support. No significant difference was found between the coarse and fine mesh.

Another important consideration is the fact that the inner conductor is cantilevered and weighs approximately 527 kg, not including any filled water cooling lines. ANSYS model results were confirmed with hand calculations and found the the bending stress on the welds from the inner conductor are about 1.9 MPa while the shear stress is about 0.84 MPa. Additionally, the welds from the inner conductor to the end ring must be considered and the bending stress is 1.9 MPa while the shear stress is 0.84 MPa. The peak von Mises stress in the 53 MHz inner cavity from the weight, thermal strain and vacuum load was found to be 152 MPa (see Fig. 5) with 0.77 mm of deflection.

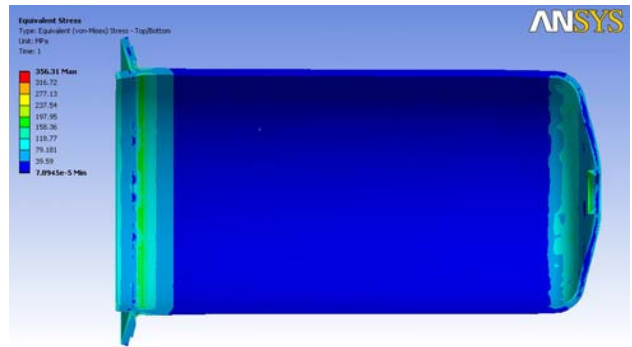

Figure 5: Peak von Mises stress for the 53 MHz cavity.

*Frequency*

The structural solution also provides the displacements of the cavity wall due to the various loading conditions. An analysis has been carried out to determine the frequency shifts associated with cavity distortion due to structural loading and external atmospheric pressure. Since the displacements are small and the frequency varies linearly with the load applied, the effects from the various loads can be evaluated separately and the results superimposed as appropriate.

A modal analysis was done in Classical ANSYS and in the vertical and horizontal direction, the frequency was found to be 48.5 Hz as shown in Fig. 6. In the axial direction, the frequency was found to 98 Hz. These frequencies are acceptable from a physics standpoint.

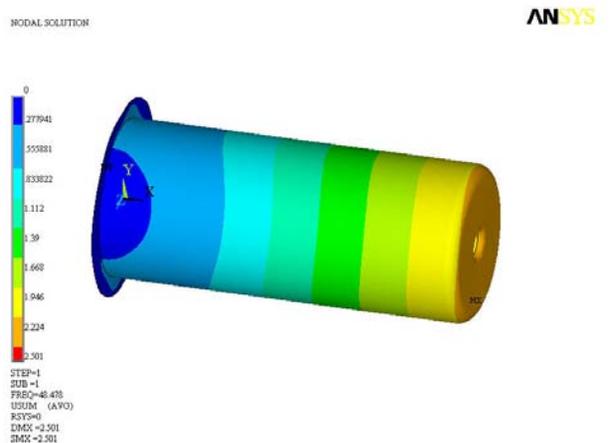

Figure 6: Fundamental frequency ANSYS result.

## CONCLUSIONS

A prototype RF cavity was built to identify various higher order modes, as well as in measuring key parameters associated with the main mode such as Q, dependency of frequency on gap spacing, and sensitivity to tuner depth and z-position [7]. The copper cavity shells have been fabricated in a forging process and purchase requisitions for the majority of the machining and other parts have been approved.

A prototype of the cavity is planned for the end of the year. Three cavities are scheduled to be installed in the Recycler Ring in early 2012. Two of the cavities will be operational; the third one will be a spare.

## ACKNOWLEDGEMENTS

The authors wish to express their appreciation to the following colleagues for their assistance in this work: Mike May, Eric Pirtle, Joseph Dey, and John Reid.

## REFERENCES

[1] I. Kourbanis, "Slip Stacking with the 53 MHz RR Cavities." NOvA-doc-4727-v1, p. 1. (2009).

[2] H. Elnaiem, "Prototype NOvA RF Cavity for the Fermilab Recycler Ring," May $18^{th}$-August $7^{th}$, 2009, internal publication, p. 1. (2009)

[3] N. Hartman and R.A. Rimmer, "Electromagnetic, Thermal, and Structural Analysis of RF Cavities using ANSYS," PAC July 2001, New York, NY (2001).

[4] ANSYS is a registered trademark of SAS IP, Inc.

[5] D. Wildman, Fermi National Accelerator Laboratory, private email communication, (October 13, 2009).

[6] W. Young, "Roarks' Formulas for Stress & Strain," McGraw-Hill, Inc, p. 429 (1989).

[7] H. Elnaiem, p. 13.